\documentstyle[emulateapj,epsfig]{article}
\slugcomment{Ap. J. (vol.573) in press}

\begin{document}

\title{Cosmic Ray Production of $^6$Li by Structure Formation Shocks in the
Early Milky Way:
       A Fossil Record of Dissipative Processes during Galaxy Formation}

\author{Takeru Ken Suzuki$^{1,2}$}
\and
\author{Susumu Inoue$^1$}
\altaffiltext{1}{Division of Theoretical Astrophysics,
                 National Astronomical Observatory,
                 2-21-1 Osawa, Mitaka, Tokyo, Japan 181-8588; 
                 stakeru@th.nao.ac.jp; inoue@th.nao.ac.jp}
\altaffiltext{2}{Department of Astronomy, Faculty of Science, University of
Tokyo,
                 7-3-1 Hongo, Bunkyo-ku, Tokyo, Japan 113-0033}

\begin{abstract}

While the abundances of Be and B observed in metal-poor halo stars
 are well explained as resulting from
 spallation of CNO-enriched cosmic rays (CRs) accelerated by supernova shocks,
 accounting for the observed $^6$Li in such stars with supernova CRs is
more problematic.
Here we propose that
 gravitational shocks induced by infalling and merging sub-Galactic clumps
 during hierarchical structure formation of the Galaxy
 should dissipate enough energy at early epochs,
 and CRs accelerated by such shocks
 can provide a natural explanation of the observed $^6$Li. 
In clear constrast to supernovae,
 structure formation shocks do not eject freshly synthesized CNO nor Fe,
 so that the only effective production channel at low metallicity
 is $\alpha-\alpha$ fusion,
 capable of generating sufficient $^6$Li
 with no accompanying Be or B and no direct correspondence with Fe.
Correlations between the $^6$Li abundance and the kinematic properties of
the halo stars
 may also be expected in this scenario.
Further, more extensive observations of $^6$Li in metal-poor halo stars,
 e.g. by the Subaru HDS or VLT/UVES,
 may offer us an invaluable fossil record
 of dissipative dynamical processes
 which occurred during the formation of our Galaxy.

\end{abstract}

\keywords{nuclear reactions, nucleosynthesis, abundances --- cosmic rays --- 
          stars: abundances ---
          Galaxy: formation --- Galaxy: halo --- Galaxy: kinematics and
dynamics}

\section{Introduction}
The light elements Li, Be and B
 are unique in that apart from $^7$Li,
 none can be synthesized appreciably
 in thermal environments such as stellar interiors, supernova envelopes
 or the standard Big Bang.
Instead, the bulk of these elements are believed
 to arise from nonthermal nuclear reactions induced by cosmic rays (CRs).
Their abundances observed in the Galactic disk,
 particularly for the isotopes $^6$Li, $^9$Be, and $^{10}$B,
 are well explained as being products
 of spallation processes in which CNO atoms in the interstellar medium (ISM)
 are broken up into LiBeB by collisions with CR protons or $\alpha$ particles
 (Reeves, Fowler \& Hoyle 1970, Meguzzi, Audouze \& Reeves 1971, Walker,
Mathews \& Viola 1985).
In the last decade, extensive observations
 of LiBeB in population II, metal poor halo stars (MPHS)
 have turned up new and unexpected results,
 spurring controversy as to what type of CR sources and production mechanisms
 were operating in the halo of the early, forming Galaxy
 (see review by Vangioni-Flam, Cass\'e \& Audouze 2000).

To date, most models of light element evolution in the early Galaxy
 have focused on strong shocks driven by supernovae (SNe) as the principal
sources of CRs.
Although the assumed CR composition, energy spectrum
 and the manner of LiBeB production vary in different models,
 a general consensus is that Be and B in MPHS
 mainly originate from the ``inverse'' spallation process,
 whereby CR CNO particles are transformed in flight into LiBeB
 by impinging on ISM H or He atoms (Duncan, Lambert \& Lemke 1992).
This can be realized if a sizable fraction of the CRs responsible for spallation
 comprise fresh, CNO-rich SN ejecta
 (e.g. Cass\'e, Lehoucq \& Vangioni-Flam 1995, Vangioni-Flam et al. 2000,
 Ramaty et al. 2000, Parizot \& Drury 1999,
 Suzuki, Yoshii \& Kajino 1999, hereafter SYK, Suzuki \& Yoshii 2001,
hereafter SY),
 as opposed to CRs injected from the average ISM (e.g. Fields \& Olive 1999).

The origin of $^6$Li in MPHS,
 which has been detected in only 3 stars so far
 (Hobbs 2000, Nissen 2000 and references therein),
 is more mysterious,
 as current models involving SN CRs face some difficulties.
A peculiar aspect of Li is that in addition to spallation,
 the fusion process of CR $\alpha$ particles with ambient He atoms
 can be effective, and should actually dominate Li production at low
metallicities.
(Note that while both $^7$Li and $^6$Li are synthesized in comparable amounts,
 the CR-produced $^7$Li component is generally overwhelmed
 by the ``Spite plateau'' from primordial nucleosynthesis
 in the metallicity range under consideration; e.g. Ryan et al. 2001.)
If the CR energy spectrum is taken
 to be a standard power-law distribution in momentum (\S 3),
 one requires a CR injection efficiency much higher than normally inferred
 to reproduce the $^6$Li observations,
 whether the CR composition is metal-enriched or not
 (Ramaty et al. 2000; SY, see their fig.2).
Faring better are some models which envision
 CR acceleration by multiple shocks inside superbubbles,
 where an additional, $\alpha$-enriched low energy CR component
 with a hard spectrum and an energy cutoff at few 100 MeV is present
 (e.g. Vangioni-Flam et al. 1999, Parizot \& Drury 1999).
However, such a CR component
 is at the moment only hypothetical,
 without any observational support.
This raises the question of whether there may have been other sources for
$^6$Li. 

This work investigates a new and more natural $^6$Li production scenario
 based on a previously unconsidered CR source:
 CRs accelerated at structure formation shocks,
 i.e. gravitational virialization shocks
 driven by the infall and merging of sub-Galactic gas clumps
 during the hierarchical build-up of structure in the early Galaxy.
Such shocks are inevitable consequences
 in the currently standard theory of hierarchical structure formation.
We show below that this picture gives a better explanation of the present data,
 and also provides a number of testable predictions for future observations
 that are quite distinct from SN CR models.
Our scenario also embodies unique and important implications
 for understanding the formation of our Galaxy.

\section{Cosmic Ray Sources in the Early Galaxy}
SNe are known to release $E_{\rm SN} \sim 10^{51}$ erg of kinetic energy in
each explosion,
 driving strong shocks into the ambient medium.
These SN shocks are favorable sites for efficient CR acceleration
 through the first order Fermi mechanism (e.g. Blandford \& Eichler 1987).
A plausible value for the injection efficiency $\xi_{\rm SN}$,
 i.e. the fraction of the SN kinetic energy imparted to CRs, is $10-20\%$,
 deduced from comparison of the SN rate
 and the energy content of CRs currently observed in the Galactic disk
 (Berezinskii et al. 1990, Chap.I, \S 4).

The global energetics of SNe for the early Galactic halo is more uncertain,
 but may be estimated as follows.
We assume that the rate of SNe in the early Galaxy (dominated by core
collapse SNe)
 was proportional to the global rate inferred from the cosmic star
formation rate
 (e.g. Madau, della Valle \& Panagia 1998).
This is $1 - 1.5$ orders of magnitude greater at redshift $z \sim 2$,
 approximately the main epoch of halo star formation, compared to $z \sim 0$;
 with a present SN rate $\sim 1/(30 - 100 {\rm yr})$,
 the average rate in the early halo is roughly $\sim 1/(3 {\rm yr})$.
Taking the duration of active star formation in the halo
 to be $\sim 5\times 10^8$ yr (\S 3),
 the total number of SNe during the halo epoch
 is $N_{\rm SN} \sim 1.5\times 10^8$,
 and its integrated energy output is
 ${\cal E}_{\rm SN} \sim N_{\rm SN}E_{\rm SN} \simeq 1.5\times 10^{59}$ erg.
If this is spread over gas with total mass $M_g \sim 3\times 10^{11} {\rm M_\odot}$,
 the average specific energy input from SNe is
 $\epsilon_{\rm SN} \sim {\cal E}_{\rm SN} \mu m_p/M_g \simeq 0.15$ keV per
particle,
 where $\mu \simeq 0.6$ is the mean molecular weight.
Alternatively, $\epsilon_{\rm SN}$ can be gauged without recourse to $M_g$
 from the total amount of heavy elements ejected by halo SNe.
Taking for example oxygen,
 its abundance at the end of the halo phase
 should correspond to [O/H] $\sim -1$ or $\sim 0.1\%$ of the total gas mass
$M_g$,
 implying a total mass in oxygen of $10^{-4} M_g$.
According to SN nucleosynthesis models (e.g. Nomoto et al. 1997),
 the oxygen yield per SN averaged over the mass distribution of progenitor stars
 is $\sim 2 M_{\odot}$.
This leads to $N_{\rm SN} \sim 5\times 10^{-5} M_g$
 and $\epsilon_{\rm SN} =N_{\rm SN} \times 10^{51} {\rm erg}/M_g \sim 0.15$
keV per particle,
 which is independent of $M_g$ and completely consistent with the above
estimate.

SNe are not the only sources of mechanical energy
 (and hence CRs through shock acceleration)
 that may have been active in the early Galaxy.
In the framework of the currently successful picture
 of hierarchical structure formation in the universe,
 large scale objects such as galaxies and clusters
 are formed through the merging and virialization of smaller subsystems,
 driven by gravitational forces acting on the dark matter.
For each merging hierarchy,
 shocks should inevitably arise in the associated baryonic gas component,
 whereby the kinetic energy of infall is dissipated
 and the gas heated to the virial temperature of the merged halo
 (e.g. White \& Rees 1978, Blumenthal et al. 1984, Peacock 1999).
It is quite plausible that
 such structure formation (SF) shocks also accelerate CRs (e.g. Miniati et
al. 2001).

A guide to how structure formation may have proceeded in our Galaxy,
 particularly for the Galactic halo,
 may be offered by the recent numerical simulations of Galaxy formation
 by Bekki \& Chiba (2000, 2001, hereafter BC; see also Steinmetz \&
M\"uller 1995).
The initial state is an overdense sphere in the expanding universe,
 superimposed on which are small scale fluctuations with a cold dark matter
(CDM) power spectrum.
Through gravitational instability,
 the perturbations on the smallest scales first develop into nonlinear
sub-Galactic clumps,
 whose typical masses are expected to be $\sim 10^6-10^7 {\rm M_{\odot}}$.
The clumps thereafter grow progressively more massive by merging with each
other,
 until by a redshift of a few, most of the mass reside in the two largest
clumps.
These then merge into a single entity in the central region
 at redshift $z \sim 2$
 (corresponding to the epoch of `formation' or `final major merger'; e.g.
Lacey \& Cole 1993),
 whereby the majority of the infall kinetic energy is virialized.
Stars eventually constituting the metal-poor halo
 are formed inside and dispersed from the merging clumps
 throughout the course of these events.

The energy dissipated at the main SF shock,
 that accompanying the major merger of the two most massive clumps,
 can be evaluated from the virial temperature of the merged system.
The total mass of this system $M_t$
 should be close to the total mass of the Galaxy $M_{t,0}$ measured today,
 as subsequent mass increase through accretion and mergers of external systems
 cannot have been very significant;
 otherwise, the thin disk, which must have formed slowly after halo formation
 and continues to exist today, would have been strongly disrupted 
 (BC, Chiba \& Beers 2001 and references therein).
The virial temperature $T_v$ for a halo of total mass $M_t$ virializing at
redshift $z$ is
 \begin{eqnarray}
 k_B T_v = {\mu m_p G M_t \over 2 r_v}
         &\simeq& 0.26 {\rm keV} \left(M_t \over 3 \times
10^{12}(h/0.7)^{-1} {\rm M_\odot}\right)^{2/3}\nonumber\\
         &\ &\times {1+z \over 3} {f(\Omega_m,\Omega_{\Lambda},z) \over
f(0.3,0.7,2)}
 \end{eqnarray}
 where $r_v$ is the virial radius,
 and $f(\Omega_m,\Omega_{\Lambda},z)$ is a factor which depends on the cosmology
 (given in e.g. Barkana \& Loeb 2001).
The numerical value corresponds to a lambda cosmology
 with $\Omega_m=0.3$, $\Omega_{\Lambda}=0.7$ and $h=0.7$.
The specific energy dissipated at the main SF shock is thus
 $\epsilon_{\rm SF} \sim 3k_B T_v/2 \simeq 0.4$ keV per particle,
 higher than the above estimate for SNe by a factor of $\simeq 2.6$,
 if $M_t = 3 \times 10^{12} {\rm M_\odot}$ is adopted.
The CR contribution from SF shocks compared to SNe
 should similarly be higher,
 as the SF shock injection efficiency $\xi_{\rm SF}$
 should not be too different from that for SNe (Miniati et al. 2001).
We note that for $M_{t,0}$ ($\simeq M_t$),
 Wilkinson \& Evans (1999) give
 $M_{t,0} = 1.9^{+3.6}_{-1.7} \times 10^{12} M_{\odot}$
 utilizing kinematic information of satellite galaxies and globular clusters,
 while Zaritsky (1999) puts a lower bound of $M_t \gtrsim 1\times 10^{12}
M_{\odot}$
 by summarizing a variety of methods.
The true value probably lies inside
 $M_t \simeq (1-5.5) \times 10^{12} M_{\odot}$,
 corresponding to a range $\epsilon_{\rm SF}/\epsilon_{\rm SN} \simeq 1.3 -
4.0$.
(The recent analysis of Sakamoto, Chiba \& Beers (2002) using a large
sample of mass tracers
 results in $M_t=2.5^{+0.5}_{-1.0} \times 10^{12} M_{\odot}$
 if the satellite Leo I is bound to the Milky Way,
 and $M_t=1.8^{+0.4}_{-0.7} \times 10^{12} M_{\odot}$ if not.)
CRs accelerated by SF shocks
 should therefore be at least as important as SN CRs, and may well dominate
at early epochs.

Another, more speculative possibility
 is outflows powered by massive black hole(s),
 which may have been active in the early Galaxy (e.g. Silk \& Rees 1998).
Our Galactic center today is known to harbor a black hole of mass
 $M_{\rm BH} \sim 3 \times 10^6 {\rm M_\odot}$ (Melia \& Falcke 2001),
 and if a fraction $\eta$ of its rest mass energy
 could have been converted into kinetic energy of a jet or a wind,
 the total ejected energy would amount to
 ${\cal E}_{\rm BH} \sim \eta M_{\rm BH} c^2
 \simeq 5.4\times 10^{59} {\rm erg} \ (\eta/0.1) (M_{\rm BH}/ 3\times 10^6
{\rm M_\odot})$.
Strong shocks induced by this outflow
 may then provide CRs energetically comparable to or greater than SNe or SF
 (c.f. Crosas \& Weisheit 1996, SY).
Black hole outflows are thus potentially interesting;
 however there are numerous ambiguities with such a picture,
 and this paper will concentrate on CRs from SF and SNe.

\section{Model Description}
SYK and SY developed a model of light element evolution based on SN CRs,
 taking into account the inhomogeneous nature of
 SN-induced chemical evolution in the early Galactic halo
 (Tsujimoto, Shigeyama \& Yoshii 1999, hereafter TSY),
 and including time-dependent calculations of CR propagation.
This model is extended
 in order to incorporate the effect of CRs from SF shocks.
Chemical evolution of the halo initiates
 with the first generation stars forming at time $t=0$,
 and lasts until $t=0.6$ Gyr
 when SN remnants have swept up all the gas in the halo
 and [Fe/H] has reached $\simeq -1.5$.

We employ assumptions and parameters deemed most plausible for the SN CRs.
The CR injection efficiency is taken to be $\xi_{\rm SN}=0.15$
 for each SN of kinetic energy $E_{\rm SN}=10^{51} {\rm erg}$.
A single power-law distribution in particle momentum
 is adopted for the CR spectrum, the form expected from standard shock
acceleration theory
 (Blandford \& Eichler 1987);
 in terms of particle energy per nucleon $E$, this is proportional to
 $(E+E_0)[E(E+2E_0)]^{-{\gamma_{\rm SN}+1 \over 2}}$,
 where $E_0=930$MeV is the nucleon rest mass energy.
The injection spectral index is chosen to be $\gamma_{\rm SN}=2.1$,
 appropriate for strong SN shocks,
 and consistent with the source spectrum inferred for present-day CRs.
As assumed in SYK and SY,
 the composition of SN CRs is a mixture
 of SN ejecta containing freshly synthesized CNO and Fe
 and the ambient ISM swept up by the SN blastwave.
This is quantified by the parameter $f_{\rm CR}$,
 the fractional mass of the ISM in the SNR shell relative to the SN ejecta mass
 that goes into CR acceleration.
Since SNe with different progenitor masses
 give varying nucleosynthetic yields (e.g. TSY),
 the composition of CRs propagating in the ISM
 also corresponds to an ensemble of SN yields.
The total injected SN CR flux at time $t$ is normalized by the SN rate at
that epoch,
 which is specified by our chemical evolution model,
 along with the above values for $\xi_{\rm SN}$ and $E_{\rm SN}$.
In order to reproduce the observed abundances of Be and B
 with our calculated results, the parameter $f_{\rm CR}$ is uniquely 
 determined to be $2\times 10^{-4}$ (see fig.2 in SY).
This then fixes the ratio of $\alpha$'s to protons in SN CRs to be $\simeq 0.14$
 (the $\alpha$/p ratio is $\simeq 0.08$ for the ISM and $\sim 0.2$ for the
SN ejecta, SY),
 which in turn fixes the resultant $^6$Li yield.

CRs from SF shocks are markedly different from SN CRs in a number of
important ways.
First and foremost, SF shocks do not synthesize fresh CNO nor Fe,
 so that the composition of these CRs
 is completely ascertained by the pre-existing ISM.
When the ISM is metal-poor,
 these shocks induce very little Be or B production through inverse spallation,
 and are only efficient at spawning Li via $\alpha-\alpha$ fusion.
Second,
 SF shocks are not necessarily strong ones,
 particularly for major mergers of systems with comparable masses (Miniati
et al. 2001),
 which applies to the case mainly envisaged here (\S 2).
Shock acceleration should then lead to 
 injection indices $\gamma_{\rm SF}$ steeper than the strong shock limit
value of 2,
 which works in favor of Li production (\S 4).
Major merger shocks may possess Mach numbers as low as $\simeq 2-3$,
 corresponding to $\gamma_{\rm SF} \simeq 2.5-3.3$;
 $\gamma_{\rm SF}=3$ is generally chosen below.
As with SN CRs,
 we take the spectral shape to be a momentum power-law distribution
 and the injection efficiency to be $\xi_{\rm SF}=0.15$.

A further distinction from SNe is that
 the SF CR flux should not entail any direct dependence on the metallicity,
 which is in fact an obstacle to predictive modeling.
While the hierarchical growth of structure
 with respect to redshift or cosmic time
 may be evaluated in concrete ways using certain formalisms
 (e.g. Lacey \& Cole 1993),
 relating this to [Fe/H] requires additional knowledge
 of how the combination of star formation, SN nucleosynthesis and chemical
evolution
 in our Galaxy proceeded with redshift or cosmic time.
This involves large uncertainties,
 and is not specified in our chemical evolution model.
As a first step,
 we choose to describe the time evolution of SF CRs in a simple,
parameterized way,
 assuming a `step function' behavior:
 SF CRs begin to be injected from a certain time $t_{\rm SF}$,
 maintains a constant flux for a duration $\tau_{\rm SF}$,
 and then returns to zero.
We take the injection duration $\tau_{\rm SF}$ to be
 roughly the dynamical time of the major merger, $\simeq 3 \times 10^8$ yr.
The injection is also assumed to be uniform,
 since the effect of the main SF shock should be global
 throughout the gas under consideration, unlike SNe.
The injected SF CR flux integrated over $\tau_{\rm SF}$
 is normalized to the above values for $\xi_{\rm SF}$ and $\epsilon_{\rm SF}$.
The true evolutionary behavior of SF shock activity relative to metallicity 
 should actually be probed through future observations of $^6$Li at low
[Fe/H] (\S 4, \S 5).

After injection by either SN or SF shocks,
 the spectral flux $F_i(E,t)$ for each CR element $i$
 evolves with time during subsequent interstellar propagation.
This is obtained from time-dependent solutions
 of the CR transport equation for a leaky box propagation model
 (Meneguzzi et al. 1971, SYK, SY),
 including escape and losses due to nuclear destruction and ionization.
At $E \sim$ 50 - 500 MeV per nucleon, the most relevant energies for LiBeB
synthesis,
 the effective loss length due to nuclear destruction $\Lambda_{{\rm
n},i}\sim 20 {\rm g cm^{-2}}$
 for CR $\alpha$'s and CNOs (Malaney \& Butler 1993),
 and the ionization loss length $\Lambda_{\rm ion} \simeq (1 - 50) (A/Z^2)
{\rm g cm^{-3}}$
 for CR particles with mass number $A$ and charge $Z$ (Northcliffe \&
Schilling 1970).

The escape length in the early halo is very uncertain,
 but we may speculate from the presumed properties
 of the gas in the merged system following the main SF shock,
 which could be of total mass $M_g \sim 3 \times 10^{11} {\rm M_\odot}$
 and spatial extent $R \sim$ 10 kpc (BC).
If the CR diffusion coefficient in this medium
 is assumed to be similar to the value inferred for the halo today,
 $D_{\rm CR} \sim 10^{29} {\rm cm^2 s^{-1}}$ (e.g. Berezinskii et al. 1990,
chap.3, \S 3),
 the escape length is roughly
 $\Lambda_{\rm esc} \sim
 (3M_g/4\pi R^3)(R^2/6D_{\rm CR})c \gtrsim 200 {\rm g cm^{-2}}$.
To be compared are $\Lambda_{{\rm n},i}$ and $\Lambda_{\rm ion}$ above,
 implying that escape is unimportant here,
 and ionization and nuclear destruction losses determine the CR spectra 
 in the energy range effective for light element production.
We set $\Lambda_{\rm esc} = 100{\rm g cm^{-2}}$ in our calculations below,
 although the exact value is irrelevant
 so long as this is larger than $\Lambda_{\rm ion}$ or $\Lambda_{{\rm n},i}$.
It is emphasized that under such conditions,
 the LiBeB production rate does not depend on the gas density $n_g$
 (as opposed to the situation considered in e.g. Fields et al. 2001):
 since both ionization and nuclear destruction loss rates are proportional
to $n_g$,
 any increase (decrease) in the density of gas containing target particles
 is always compensated by a decrease (increase)
 in the propagated CR flux due to increased (decreased) losses.

Using the transported spectra,
 we calculate the CR production of LiBeB in the ISM 
 including all three types of reactions described in $\S 1$:
 forward spallation of ISM CNO by CR protons and $\alpha$'s,
 inverse spallation of CR CNOs by ISM H and He,
 and the fusion of CR $\alpha$'s with ISM He.
For more details, consult SYK and SY.
We do not consider here
 the potentially complicating effects of stellar depletion,
 which are highly uncertain at the moment (e.g. Pinsonneault et al. 1999)
 and await further studies.

\section{Results and Discussion}
\begin{figure}
 \figurenum{1} 
 \epsscale{1.0} 
 \plotone{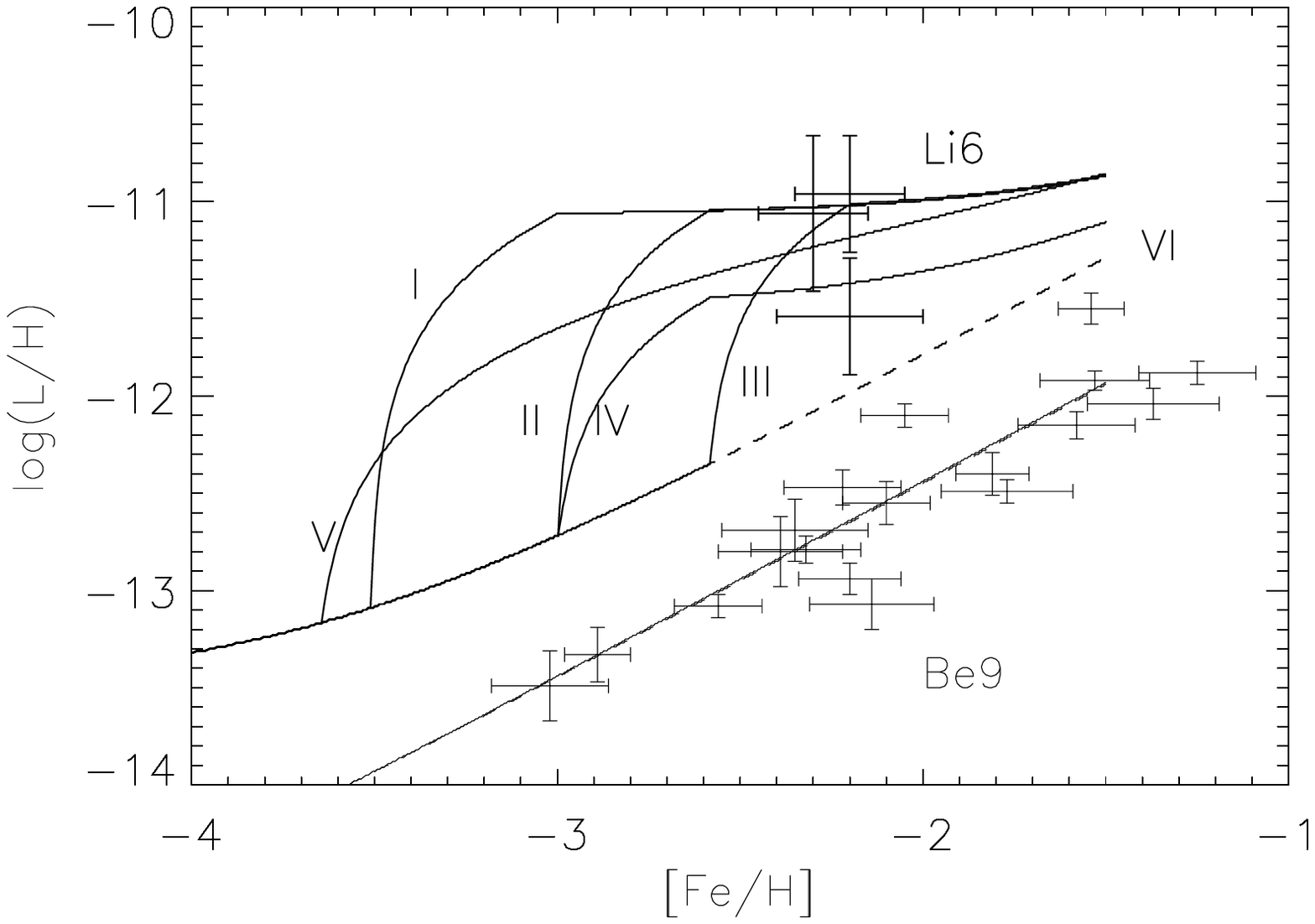} 
 \caption{
Model results of $^6$Li/H (thick) and Be/H (thin) vs. [Fe/H],
 for SN CRs only (dashed curves) and SN plus SF CRs (solid curves),
 each label corresponding to the parameter set described in the text.
Also plotted are current observational data
 for $^6$Li (thick markers) and Be (thin markers).
} 
 \label{fig:libe}
 \end{figure}

For our calculations, we have selected the following sets of parameters
 for $t_{\rm SF}$, $\tau_{\rm SF}$ and $\gamma_{\rm SF}$, respectively,
 labeled models I - V:
 I (0.12, 0.1, 3),
 II (0.22, 0.1, 3),
 III (0.32, 0.1, 3),
 IV (0.22, 0.1, 2) and
 V (0.1, 0.5, 3), where $t_{\rm SF}$ and $\tau_{\rm SF}$ are in units of Gyr;
 model VI is the case of SN CRs only.
These were chosen to provide results
 exemplary of light element production by SF CRs, in contradistinction to
that by SN CRs.
The evolution of $^6$Li and Be vs. metallicity
 calculated for each model until the end of halo chemical evolution
([Fe/H]$\simeq -1.5$)
 is shown in Fig.\ref{fig:libe},
 along with the current observational data for $^6$Li
 (Smith et al. 1998, Cayrel et al. 1999, Nissen et al. 2000)
 and Be (Boesgaard et al. 1999) in MPHS.

We discuss some salient points regarding these results.
First, it is confirmed that with our fiducial parameters,
 production by SN CRs alone (VI) 
 works very well for the observed Be (and B, not shown),
 yet falls short of the observed $^6$Li.
Accounting for this by SN CRs demands a much larger value
 of $E_{\rm SN}$, $\xi_{\rm SN}$ and/or $\gamma_{\rm SN}$,
 which must be supplemented with a larger $f_{\rm CR}$
 in order for Be and B to be consistently reproduced together (see fig.2 in SY).

In contrast, with reasonable values for
 $\epsilon_{\rm SF}$, $\xi_{\rm SF}$ and $\gamma_{\rm SF}$,
 production by SF CRs is capable of explaining the current $^6$Li data
quite adequately.
This mainly owes to two facts:
 1) SF shocks are more energetic than (or at least as energetic as)
 SN shocks, as estimated in \S 2, and
 2) SF CRs can generate $^6$Li at early epochs independently of the metallicity.
Regardless of the early evolutionary behavior,
 identical $^6$Li abundances are attained at the end of the halo phase
 for a given $\gamma_{\rm SF}$ (I - III, V),
 since this is determined by the time-integrated CR flux,
 for which we had assumed a fixed value.
Compared to a flat spectral index of $\gamma_{\rm SF}=2$ (IV),
 a steeper one of $\gamma_{\rm SF}=3$, more appropriate for low Mach number
SF shocks (\S 3),
 results in a larger $^6$Li yield, by about factor of 3.
This is because with a constant total CR energy,
 a steeper index implies a larger CR flux
 in the subrelativistic energy range $E \sim 10$ -- $100$ MeV,
 where the $^6$Li production cross section peaks (see fig.1 in SY).
Note that a conservatively lower specific energy for SF shocks,
 $\epsilon_{\rm SF} \sim 0.15$ keV/particle (i.e. comparable to SNe),
 can still be consistent with the available $^6$Li data
 provided that $\gamma_{\rm SF} \simeq 3$.

As already stressed, 2) is a consequence of
 $^6$Li synthesis being dominated by $\alpha-\alpha$ fusion,
 and SF shocks not creating any new Fe.
Depending on the onset time and duration of the SF shock,
 the $^6$Li abundance may potentially reach large values
 quickly at very low metallicity,
 which can be followed by a plateau or a very slow rise.
On the other hand, SNe unavoidably give forth to freshly synthesized Fe,
 so a correlation with Fe/H must arise;
 in fact $^6$Li/H vs. Fe/H for SN CRs can never be much flatter than linear
 (e.g. Vangioni-Flam et al. 1999, Parizot \& Drury 1999, Fields \& Olive
1999, Ramaty et al. 2000).
Moreover, since SF shocks do not eject fresh CNO,
 they produce very little Be or B through spallation;
 only a minuscule contribution can appear as the ISM becomes metal-enriched.
This may allow an extremely large $^6$Li/Be ratio at low [Fe/H].
Conversely, we see that SN CRs must play an indispensable role
 in generating the Be and B observed in MPHS.

Independent of the particular evolutionary parameters,
 the following abundance trends are characteristic of SF CR production
 and should serve as distinguishing properties of the scenario for future
observations.
Going from high to low metallicity:
 a plateau or a very slow decrease in $\log ^6$Li/H vs. [Fe/H],
 followed by a steeper decline in some range of [Fe/H]
 corresponding to the main epoch of SF;
 a steady increase in $^6$Li/Be, possibly up to values exceeding $\simeq 100$,
 also followed by a downturn.
These traits are very distinctive and not expected in SN CR models,
 for which the slope of $\log ^6$Li/H - [Fe/H] must be $\simeq 1$ or greater,
 and the $^6$Li/Be ratio constant at sufficiently low [Fe/H].
Distinction from any production processes in the early universe (e.g.
Jedamzik 2000)
 should also be straightforward,
 as they predict a true plateau down to the lowest [Fe/H],
 in contrast to an eventual decrease for SF CR models.
Further, unique diagnostic features are discussed in \S 5.

Observing $^6$Li in MPHS has proven to be difficult in the past,
 as measurement of its weak isotopic shift feature
 relative to the much stronger $^7$Li line
 requires spectroscopy with very high resolution and signal to noise.
The present database of only 3 positive detections
 in a narrow range of [Fe/H] ($\sim$ -2.5 -- -2)
 plus a number of upper limits
 (Hobbs 2000, Nissen 2000 and references therein)
 is obviously insufficient
 for distinction between our SF CR model and various other models.
However, the new generation of large aperture telescopes
 equipped with high resolution spectrographs,
 e.g. the Subaru HDS or VLT/UVES,
 should in the near future bring about
 a larger sample of MPHS with accurately determined $^6$Li/H and $^6$Li/Be
 over a wide range of [Fe/H],
 and greatly help us toward deciphering the true origin of $^6$Li in MPHS.

\section{$^6$Li as Fossil Record of Dissipative Processes during Galaxy
Formation}

A truly unique and intriguing aspect of the SF CR picture is that
 $^6$Li in MPHS can be interpreted and utilized as a fossil record
 of dissipative gas dynamical processes in the early Galaxy.
Of particular interest are various correlations expected
 between the $^6$Li abundance and the kinematic properties of the stars.
On the one hand,
 $^6$Li arises as a consequence of gaseous dissipation through
gravitational shocks,
 and survives to this day as signatures
 of the dynamical history of hierarchical structure formation in the early
Galaxy.
On the other,
 the kinematic characteristics of stars presently observed should reflect
 the past dynamical state of their parent gas systems,
 because once stars form, they become collisionless
 and have long timescales for phase space mixing (e.g. Chiba \& Beers 2000).
Interesting relationships may then exist among the two observables.

For example,
 recent intensive studies of the structure and kinematics of MPHS in our Galaxy
 based on Hipparcos data (e.g. Chiba \& Beers 2000, 2001)
 have elucidated the detailed characteristics of our Galaxy's halo,
 such as its two-component nature:
 an inner halo which is flattened and rotating,
 and an outer halo which is spherical and non-rotating.
This dichotomy has been suggested to result from
 differences in the physical processes responsible for their formation,
 dissipative gas dynamics being crucial for the former,
 and dissipationless stellar dynamics determining the latter.
The numerical simulations of Galaxy formation by BC support this conjecture
 (see \S 2 for a rough outline):
 the outer halo forms through dissipationless merging of small sub-Galactic
clumps
 that have already turned into stars (c.f. Searle \& Zinn, 1978),
 whereas the inner halo mainly forms through dissipative merging and
accretion of larger clumps
 that are still gas rich (c.f. Eggen, Lynden-Bell \& Sandage, 1962).
In our scenario,
 $^6$Li production is a direct outcome
 of the principal gas dissipation mechanism of gravitational shock heating.
If the above inferences on the formation of halo structure are correct,
 $^6$Li should be systematically more abundant
 in stars belonging to the inner halo compared to those of the outer halo,
 a clearly testable prediction.
An important prospect is that
 $^6$Li may provide a quantitative measure
 of the effectiveness of gas dynamical processes during formation of halo
structure,
 rather than just the qualitative deductions allowed by kinematic studies.

Another possibility regards the main epoch of SF with respect to
metallicity (\S 3).
As already mentioned,
 the relation between $^6$Li/H and Fe/H
 should mirror the time evolution of dissipative energy release through SF
shocks,
 but is complicated by being convolved with
 the uncertain ingredients of star formation and chemical evolution.
The Fe abundance can be a bad tracer of time, especially at low [Fe/H]
 where the effects of dispersion in SN yields can be extremely large (TSY, SY).
Stellar kinematics information may offer a handle on this problem.
The observed relation between [Fe/H] and $<V_\phi>$, the mean azimuthal
rotation velocity of MPHS,
 seems to manifest a distinctive kink around [Fe/H] $\sim$ -2 (Chiba \&
Beers 2000).
Through chemo-dynamical modeling of the early Galaxy,
 BC have proposed that this kink may correspond to the epoch of the major
merger (\S 3).
If this was true,
 a simple expectation in the context of the SF CR model
 is that $^6$Li/H should be just rising near this value of [Fe/H],
 which is the range occupied by the currently $^6$Li-detected stars;
 also expected are a steep decline at lower [Fe/H],
 as well as a plateau or slow rise at higher [Fe/H] (i.e. close to model
III in fig.\ref{fig:libe}).
However, any inferences related to [Fe/H]
 are always subject to the chemical evolution ambiguities.
A more reliable and quantitative answer may be achieved
 by looking for correlations between $^6$Li/H and $<V_\phi>$ without
recourse to [Fe/H],
 as $^6$Li is a direct and pure indicator of dynamical evolution in the
early Galaxy.

Thus the SF CR model for $^6$Li bears important implications
 for understanding how our Galaxy formed.
If the above mentioned trends are indeed observed,
 it would not only confirm the structure formation origin of $^6$Li,
 but may potentially point to new studies of ``$^6$Li Galactic archaeology'',
 whereby extensive observations of $^6$Li in MPHS
 can be exploited as a robust and clear-cut probe
 of dissipative dynamical processes that were essential for the formation
of the Galaxy.

\section{Summary}

We have put forth a new scenario
 for the currently puzzling origin of $^6$Li observed in metal-poor halo stars:
 production by cosmic rays accelerated at structure formation shocks,
 driven by the hierarchical infall and merging of sub-Galactic structure
 during the formation of our Galaxy.
Several predictions
 are quite distinct from models involving supernova cosmic rays
 and clearly testable in the near future,
 such as the behavior of $^6$Li/H vs. [Fe/H] and $^6$Li/Be vs. [Fe/H] at
low [Fe/H],
 as well as possible correlations between $^6$Li/H and the kinematic
properties of halo stars.
Since $^6$Li can be construed as a fossil record
 of dissipative processes during Galaxy formation,
 further observations of this isotope in halo stars may offer us
 a unique and invaluable insight into the past dynamical history of our Galaxy.

\acknowledgments

The authors are very grateful to M. Chiba for extensive and fruitful
discussions.
We also thank G. Mathews, T. Kajino and M. Nagashima for valuable conversations,
 Y. Yoshii for assistance with our code, and the anonymous referee for
helpful comments.
T.K.S. is supported by the JSPS Research Fellowship for Young Scientists,
grant 5936.

\clearpage

\end{document}